\documentclass[acmsmall,screen]{acmart}
\usepackage{graphicx} 

\setcopyright{none}
\copyrightyear{2026}
\acmYear{2026}
\acmDOI{}

\usepackage{fancyhdr}
\usepackage{natbib}
\usepackage{mdwtab}
\usepackage{hyperref}
\usepackage{hypcap}
\usepackage[capitalize,nameinlink]{cleveref}
\usepackage{textcomp}
\usepackage{stmaryrd}
\usepackage{mathpartir}
\usepackage{amsmath}
\newtheorem{theorem}{Theorem}
\usepackage{enumitem}
\usepackage{xcolor}
\usepackage{subcaption}

\usepackage{tikz}
\usetikzlibrary{matrix}

\usepackage{listings}
\lstdefinestyle{scheme}{
  language=Lisp,
  basicstyle=\ttfamily,
  upquote=true,
  columns=flexible,
  mathescape=true,
  breaklines=true,
  morekeywords={defrel,fresh,conj,disj,factor},
  deletekeywords={sum,let,map,cons,null,and,or,cond,equal},
  keepspaces=true,
}
\usepackage{ebproof}

\title[All for one and none forall: Compiling polymorphic relations without monomorphization]{All for one and none forall:\linebreak[3]Compiling polymorphic relations without monomorphization}
\author{Dmitri Volkov}
\email{dvolkov@iu.edu}
\author{Yafei Yang}
\email{yafyang@iu.edu}
\author{Chung-chieh Shan}
\email{ccshan@iu.edu}
\affiliation{
  \institution{Indiana University}
  \city{Bloomington}
  \country{USA}}

\newcommand{\den}[1]{\llbracket #1 \rrbracket}

\newcommand{\type}{\mathrel{\mathbf{type}}}
\newcommand{\goal}{\mathrel{\mathbf{goal}}}
\newcommand{\program}{\mathrel{\mathbf{program}}}
\newcommand{\query}{\mathrel{\mathbf{query}}}

\DeclareMathOperator{\shell}{\mathit{shell}}
\DeclareMathOperator{\holes}{\mathit{holes}}
\DeclareMathOperator{\envshell}{\mathit{envshell}}
\DeclareMathOperator{\envholes}{\mathit{envholes}}

\newcommand{\eqpat}{\leftrightharpoons}
\newcommand{\eqpatD}{\eqpat_{\Delta}}

\DeclareMathOperator{\enforceeqpatD}{\mathit{enforce}_{\eqpatD}}
\DeclareMathOperator{\compile}{\mathit{compile}}

\begin{document}

\settopmatter{printacmref=false}
\settopmatter{printfolios=true}
\renewcommand\footnotetextcopyrightpermission[1]{}
\pagestyle{fancy}
\fancyfoot{}
\fancyfoot[R]{miniKanren'26}
\fancypagestyle{firstfancy}{
  \fancyhead{}
  \fancyhead[R]{miniKanren'26}
  \fancyfoot{}
}
\makeatletter
\let\@authorsaddresses\@empty
\makeatother

\begin{abstract}
    We present a new approach for implementing polymorphism for bottom-up relational languages that avoids monomorphization.
    We begin by introducing semiringKanren, a bottom-up weighted relational programming language.
    We extend this base language to support polymorphism.
    We describe a new method to compile polymorphic semiringKanren programs into non-polymorphic ones, based on \emph{equality patterns} and \emph{large-enough instances} of polymorphic relations.
    We explain the correctness of this method.
    Finally, we consider existing work and suggest directions for future research.
\end{abstract}

\maketitle
\thispagestyle{firstfancy}

\section{Introduction}

Relational programming languages (such as Prolog, miniKanren, datalog) are usually implemented via top-down search or bottom-up fact collection.
Top-down relational languages can effectively handle recursive or untyped data, but are also prone to getting stuck searching ineffectual branches.
Bottom-up relational languages generally cannot handle arbitrary recursive data and are less space-efficient, but have a clearer semantics and avoid many of the traps of top-down languages.

We use the term \emph{polymorphism} to describe code that can operate on different types of data.
Polymorphism allows programmers to use a single piece of code on different types of data, which reduces the need to manually reimplement functionality.
In this work, we focus on \emph{parametric} polymorphism, where unknown types are represented with \emph{type variables} and arbitrary operations are not allowed on values of unknown types.

In particular, we show how to support parametric polymorphism in a bottom-up relational programming language.
This may feel paradoxical:
how can we collect facts of the form ``$x$ is related to $y$'' when we do not know which domains $x$ and $y$ are drawn from?
One common approach is to track which concrete types are used for $x$ and $y$ in practice, and generate code specialized for these types.
This approach is known as \emph{monomorphization}.
Unfortunately, monomorphization can be space-inefficient and result in long compile times.
Furthermore, monomorphization does not generalize; what if we want to call a relation with new types that do not have a specialized instance?
In particular, monomorphization does not support \emph{polymorphic recursion}, where polymorphic code may call itself at a different type than it was originally called.

In this work, we extend the semiringKanren relational programming language \cite{volkovCommittingBitRelational2025a} to support polymorphic relations.
In particular, we introduce a method for compiling polymorphic programs into non-polymorphic programs
that largely avoids monomorphization.
We show how \emph{equality patterns} can be used to extract the underlying structure of
\emph{large-enough} instances of polymorphic relations to handle calls at different types.

\section{Overview}\label{section:overview}

We introduce semiringKanren and our compilation approach by example.

\subsection{Equal}

Consider the relation:
\begin{lstlisting}[mathescape]
(defrel (equal (x : (Prod (Sum Unit Unit) (Sum Unit Unit)))
               (y : (Prod (Sum Unit Unit) (Sum Unit Unit))))
  (== x y))
\end{lstlisting}
This relation takes two arguments (\lstinline{x} and~\lstinline{y}) and asserts that they are equal.
In semiringKanren, values inhabit \emph{algebraic data types}.
We use the notation \lstinline[mathescape]{$v$:$\tau$} to say the value $v$ inhabits the type $\tau$.
Values and their inhabited algebraic data types include:
\begin{itemize}
    \item \lstinline{sole}, which inhabits \lstinline{Unit}.
    \item \lstinline[mathescape]{(left $v_1$)}, which inhabits \lstinline[mathescape]{(Sum $\tau_1$ $\tau_2$)}
    when \lstinline[mathescape]{$v_1$:$\tau_1$}.
    \item \lstinline[mathescape]{(right $v_2$)}, which inhabits \lstinline[mathescape]{(Sum $\tau_1$ $\tau_2$)}
    when \lstinline[mathescape]{$v_2$:$\tau_2$}.
    \item \lstinline[mathescape]{(pair $v_1$ $v_2$)} which inhabits \lstinline[mathescape]{(Prod $\tau_1$ $\tau_2$)}
    when \lstinline[mathescape]{$v_1$:$\tau_1$} and \lstinline[mathescape]{$v_2$:$\tau_2$}.
\end{itemize}
Hence, the type \lstinline{(Prod (Sum Unit Unit) (Sum Unit Unit))} is inhabited by four values:
\begin{enumerate}[start=0]
    \item \lstinline{(pair (left sole) (left sole))}
    \item \lstinline{(pair (left sole) (right sole))}
    \item \lstinline{(pair (right sole) (left sole))}
    \item \lstinline{(pair (right sole) (right sole))}
\end{enumerate}
We say the \emph{size of the type}, notated $|\text{\lstinline{(Pair (Sum Unit Unit) (Sum Unit Unit))}}|$, is $4$.
We can use the numbers above as indices into an array tracking success/failure for each value taken by \lstinline{x} and \lstinline{y}:
\begin{equation}
\label{e:identity4}
    \begin{matrix}
    & y \rightarrow \\
    x \downarrow &
    \begin{bmatrix}
        1 & 0 & 0 & 0 \\
        0 & 1 & 0 & 0 \\
        0 & 0 & 1 & 0 \\
        0 & 0 & 0 & 1
    \end{bmatrix}
    \end{matrix}
\end{equation}
As we can see, the relation succeeds when the values of \lstinline{x} and \lstinline{y} are equal, and fails otherwise.
This has the structure of an identity matrix.

\subsection{Polymorphic Equal}

Our previous definition of \lstinline{equal} works with only one specific type.
We generalize it here to operate on unknown types, represented by \emph{type variables}.
\begin{lstlisting}[mathescape]
(defrel (equal (x : $\alpha$) (y : $\alpha$))
  (== x y))
\end{lstlisting}
Here, \lstinline{x} and \lstinline{y} must inhabit the same unknown type, represented by the type variable $\alpha$.
When called at the type \lstinline{(Prod (Sum Unit Unit) (Sum Unit Unit))}, the \lstinline{equal} relation is represented by the array~\eqref{e:identity4} above.
When \lstinline{equal} is instead called at a smaller type such as \lstinline{(Sum Unit Unit)}, which is inhabited by only two values, it is represented by the smaller array:
\begin{equation}
\label{e:identity2}
    \begin{bmatrix}
        1 & 0 \\
        0 & 1
    \end{bmatrix}
\end{equation}
No matter the type~$\alpha$ that is used, we represent \lstinline{equal} with some sort of identity matrix.
Intuitively, it should be possible to represent this structure in a compact way.
As we will show, the two-by-two identity matrix~\eqref{e:identity2} can represent \lstinline{equal} for all types!

\subsection{Sum-Swap}

Now consider the relation:
\begin{lstlisting}[mathescape]
(defrel (sum-swap (x : (Sum $\alpha$ $\beta$)) (y : (Sum $\beta$ $\alpha$)))
  (disj
    (fresh ((a : $\alpha$))
      (conj (== x (left a)) (== y (right a))))
    (fresh ((b : $\beta$))
      (conj (== x (right b)) (== y (left b))))))
\end{lstlisting}
This relation replaces top-level \lstinline{left}s in a sum type with \lstinline{right}s, and vice versa.
We introduce three new constructs here: \lstinline{fresh}, \lstinline{disj}, and \lstinline{conj}.
These constructs, along with \lstinline{==}, are called \emph{goals}.
The \lstinline{fresh} goal introduces a new variable in scope, before evaluating subgoals.
The \lstinline{disj} goal succeeds when any of its subgoals succeeds,
and the \lstinline{conj} goal succeeds when all of its subgoals succeed.

When called at $\alpha=\text{\lstinline{Unit}}$ and $\beta=\text{\lstinline{(Sum Unit Unit)}}$,
the call
\begin{lstlisting}
(sum-swap (left sole) (right sole))
\end{lstlisting}
succeeds,
whereas the call
\begin{lstlisting}
(sum-swap (right (left sole)) (right sole))
\end{lstlisting}fails.
As before, we can represent this relation as a matrix:
\begin{equation}
    \begin{bmatrix}
        0 & 0 & 1 \\
        1 & 0 & 0 \\
        0 & 1 & 0
    \end{bmatrix}
\end{equation}
If the same polymorphic relation is called at larger types where $|\alpha|=|\beta|=3$, we get the matrix:
\begin{equation}
    \begin{bmatrix}
        0 & 0 & 0 & 1 & 0 & 0 \\
        0 & 0 & 0 & 0 & 1 & 0 \\
        0 & 0 & 0 & 0 & 0 & 1 \\
        1 & 0 & 0 & 0 & 0 & 0 \\
        0 & 1 & 0 & 0 & 0 & 0 \\
        0 & 0 & 1 & 0 & 0 & 0
    \end{bmatrix}
\end{equation}
Again, both matrices have similar structures: identity sub-matrices in the top right and bottom left corners.

\subsection{Two-Valued}

Consider the following relation:
\begin{lstlisting}[mathescape]
(defrel (two-valued (x : $\alpha$))
  (fresh ((y : $\alpha$))
    (=/= x y)))
\end{lstlisting}
Here, we use the \lstinline{=/=} subgoal to assert that the fresh variable \lstinline{y} is different from the relation argument~\lstinline{x}.
When $\alpha=\text{\lstinline{(Sum Unit Unit)}}$,
the call \lstinline{(two-valued (left sole))} succeeds because \lstinline{y} can be \lstinline{(right sole)}.
Symmetrically, \lstinline{(two-valued (right sole))} also succeeds.
Thus for $|\alpha|=2$, the matrix for \lstinline{two-valued} is:
\begin{equation}
    \begin{bmatrix}
        1 & 1
    \end{bmatrix}
\end{equation}
However, when $\alpha=\text{\lstinline{Unit}}$, the only value \lstinline{x} and \lstinline{y} can take is \lstinline{sole}.
Thus the call \lstinline{(two-valued sole)} fails, and is represented by the matrix:
\begin{equation}
    \begin{bmatrix}
        0
    \end{bmatrix}
\end{equation}
Unlike in the previous examples, these two matrices do not have similar structures.

\subsection{Implementation Without Monomorphization}

Our overall implementation strategy is as follows.
First we compute for each relation a finite set of monomorphic instances that are \emph{large enough} to handle most calls.
Then we compile each call to convert between the caller's larger types and the callee's smaller types by enforcing an \emph{equality pattern} upon the arguments.
For relation calls at types smaller than those used by the large-enough instances, we generate specialized monomorphic instances.
The result of compilation is a monomorphic program that can be run as usual.

\section{The Language}

We formalize our language in this section.

\subsection{Syntax}

We give the syntax of polymorphic semiringKanren in \cref{fig:poly-syntax}.
Like semiringKanren \cite{volkovCommittingBitRelational2025a}, polymorphic semiringKanren supports \lstinline{Unit}, \lstinline{Pair}, and \lstinline{Sum} types and their values.
There are two main differences.

First, polymorphic semiringKanren supports type variables.
These type variables are introduced by relation definitions (\lstinline{defrel}).
Unlike with the examples given in \cref{section:overview}, we omit the syntactic sugar and require that each relation definition introduces all the type variables it uses.
In a relation definition \texttt{(defrel ($R$ $\forall \alpha \dots$ . ($x:\tau$) $\dots$) $g$)}, we first introduce type variables $\alpha \dots$.
Then each logic variable of the relation is assigned a type, which can contain these type variables.
Within the body $g$ of the relation, the types of logic variables introduced by a \lstinline{fresh} can also contain these type variables.
Logic variables introduced by a query \texttt{(run (($x : \tau$) $\ldots$) g)} cannot use type variables.

Second, polymorphic semiringKanren relaxes the restriction on what arguments can be passed to relations.
In monomorphic semiringKanren, arguments passed to \lstinline{==}, \lstinline{=/=}, and relations must be logic variables, so we cannot write a goal like
\begin{lstlisting}
(== $x$ (pair (left$_{\texttt{(Sum Unit Unit)}}$ $y$) $z$))
\end{lstlisting}
Instead, monomorphic semiringKanren deals with \lstinline{Unit}, \lstinline{Pair}, and \lstinline{Sum} types using the goal constructs \lstinline{soleo}, \lstinline{pairo}, \lstinline{lefto}, and \lstinline{righto}.
The arguments passed to these constructs must also be logic variables.
Thus, the \lstinline{==} goal above needs to be expressed in monomorphic semiringKanren as
\begin{lstlisting}
(fresh (($w$ : (Sum Unit Unit)))
  (conj (lefto $w$ $y$) (pairo $x$ $w$ $z$)))
\end{lstlisting}
In polymorphic semiringKanren, an argument to a relation can be any value constructed from logic variables using \lstinline{sole}, \lstinline{pair}, \lstinline{left}, and \lstinline{right}.
Therefore, we don't need the goal constructs \lstinline{soleo}, \lstinline{pairo}, \lstinline{lefto}, and \lstinline{righto}.
This makes it slightly easier to express the same goal in polymorphic semiringKanren, though the two ways can be converted to each other.

Our grammar includes \emph{monomorphized relation variables}.
These are used in our denotational semantics, which is described in \cref{section:semantics}.

The \lstinline{factor} goal is standard for semiringKanren and confers weight drawn from a semiring to a program branch.
Traditional relational programming can be thought of operating over the boolean semiring; failure is $\bot$, and success is $\top$.
This generalizes to arbitrary semirings, where a zero value denotes failure and nonzero values denote success.
The \lstinline{conj} goal can be thought of as semiring multiplication,
and the \lstinline{disj} goal can be thought of as semiring addition.
While weighted relational programming is not the focus of this paper, polymorphic semiringKanren does support weighted relational programs, with the restriction that semiring addition be idempotent: $1 + 1 = 1$ (or $\top \vee \top = \top$).
We explain this in more detail in section \ref{sec:eqpat-preserves-weight}.
Further information about weighted relational programming in semiringKanren is available in \cite{volkovCommittingBitRelational2025a}.

\begin{figure}[t!]
\[
\begin{array}{lrcl}
    \textrm{Programs} & P & ::= & \mathit{Rel}\dots\mathit{Q} \\
    \textrm{Relations} & \mathit{Rel} & ::= & \texttt{(defrel (\(R\) \(\forall \alpha \dots\) . (\(x:\tau\)) \(\dots\)) \(g\)) } \\
    \textrm{Queries} & Q & ::= & \texttt{(run ((}x:\tau\texttt{)}\:\ldots\texttt{) }g\texttt{)}\\
    \textrm{Goals} & g & ::= & \texttt{(conj \(g\) \(g\))} \\
    & & | & \texttt{(disj \(g\) \(g\))} \\
    & & | & \texttt{(fresh ((\(x:\tau\)) \(\dots\)) \(g\))} \\
    & & | & \texttt{(== \(v\) \(v\))} \\
    & & | & \texttt{(=/= \(v\) \(v\))} \\
    & & | & \texttt{(\(R\) \(v\) \(\dots\))} \\
    & & | & \texttt{(factor \(r\))} \\
    \textrm{Values} & v & ::= & x \\
    & & | & \texttt{sole} \\
    & & | & \texttt{(pair \(v\) \(v\))} \\
    & & | & \texttt{(left\(_{\texttt{(Sum \(\tau_1\) \(\tau_2\))}}\) \(v\))} \\
    & & | & \texttt{(right\(_{\texttt{(Sum \(\tau_1\) \(\tau_2\))}}\) \(v\))} \\
    \textrm{Types} & \tau & ::= & \texttt{Unit} \\
    & & | & \texttt{(Prod \(\tau_1\) \(\tau_2\))} \\
    & & | & \texttt{(Sum \(\tau_1\) \(\tau_2\))} \\
    & & | & \alpha \\
    \textrm{Relation variables} & R & & \\
    \textrm{Monomorphized relation variables} & R_{\sigma} & & \\
    \textrm{Logic variables} & x,y,z & & \\
    \textrm{Type variables} & \alpha,\beta & & \\
    \textrm{Weights} & r & \in & \mathbb{K}
\end{array}
\]
\caption{The syntax of polymorphic semiringKanren}
\label{fig:poly-syntax}
\end{figure}

\subsection{Typing}

\begin{figure}[!htbp]
\begin{subfigure}{\textwidth}
\[
  \begin{aligned}
    &\textrm{Value typing contexts} & \Delta &::= \varnothing \quad|\quad \Delta,\alpha:* \quad|\quad \Delta,x : \tau\\
    &\textrm{Relation typing contexts} & \Gamma &::= \varnothing \quad|\quad \Gamma, (R : \forall \vec\alpha . \vec\tau \rightarrow)\\
    &\textrm{Substitutions} & \sigma &::= \varnothing \quad|\quad \sigma,\alpha \mapsto \tau
  \end{aligned}
\]
\begin{mathpar}
\end{mathpar}
\vspace{-15pt}
\caption{Typing contexts}
\label{fig:poly-contexts}
\end{subfigure}
\smallskip

\begin{subfigure}{\textwidth}
\begin{mathpar}
    \infer{ }{\Delta \vdash \texttt{Unit} \type}
    \and
    \infer{
        \Delta \vdash \tau_1 \type \and \Delta \vdash \tau_2 \type
    }{
        \Delta \vdash \texttt{(Prod \(\tau_1\) \(\tau_2\))} \type
    }
    \and
    \infer{
        \Delta \vdash \tau_1 \type \and \Delta \vdash \tau_2 \type
    }{
        \Delta \vdash \texttt{(Sum \(\tau_1\) \(\tau_2\))} \type
    }
    \and
    \infer{ }{\Delta,\alpha : * \vdash \alpha \type}
\end{mathpar}
\vspace{-5pt}
\caption{Type validity $\Delta \vdash \tau \type$}
\label{fig:poly-types-valid}
\end{subfigure}
\smallskip

\begin{subfigure}{\textwidth}
\begin{mathpar}
    \infer{
    }
    {
        \Delta, x : \tau \vdash x : \tau
    }
    \and
    \infer{ }{
        \Delta \vdash \texttt{sole} : \texttt{Unit}
    }
    \and
    \infer{
        \Delta \vdash v_1 : \tau_1
        \and
        \Delta \vdash v_2 : \tau_2
    }
    {
        \Delta \vdash \texttt{(pair \(v_1\) \(v_2\))} : \texttt{(Prod \(\tau_1\) \(\tau_2\))}
    }
    \and
    \infer{
        \Delta \vdash v : \tau_1
    }
    {
        \Delta \vdash \texttt{(left\(_{\texttt{(Sum \(\tau_1\) \(\tau_2\))}}\) \(v\))} : \texttt{(Sum \(\tau_1\) \(\tau_2\))}
    }
    \and
    \infer{
        \Delta \vdash v : \tau_2
    }
    {
        \Delta \vdash \texttt{(right\(_{\texttt{(Sum \(\tau_1\) \(\tau_2\))}}\) \(v\))} : \texttt{(Sum \(\tau_1\) \(\tau_2\))}
    }
\end{mathpar}
\vspace{-5pt}
\caption{Value typing $\Delta \vdash v : \tau$}
\label{fig:poly-value-types}
\end{subfigure}
\smallskip

\begin{subfigure}{\textwidth}
\begin{mathpar}
    \infer{
        \Gamma;\Delta \vdash g_1 \goal
        \and
        \Gamma;\Delta \vdash g_2 \goal
    }{
        \Gamma;\Delta \vdash \texttt{(conj \(g_1\) \(g_2\))} \goal
    }
    \and
    \infer{
        \Gamma;\Delta \vdash g_1 \goal
        \and
        \Gamma;\Delta \vdash g_2 \goal
    }{
        \Gamma;\Delta \vdash \texttt{(disj \(g_1\) \(g_2\))} \goal
    }
    \and
    \infer{
        \Gamma;\Delta,x:\tau \vdash g \goal
    }{
        \Gamma;\Delta \vdash \texttt{(fresh ((\(x:\tau\))) \(g\))} \goal
    }
    \and
    \infer{
        \Delta \vdash v_1 : \tau
        \and
        \Delta \vdash v_2 : \tau
    }{
        \Gamma; \Delta \vdash \texttt{(== \(v_1\) \(v_2\))} \goal
    }
    \and
    \infer{
        \Delta \vdash v_1 : \tau
        \and
        \Delta \vdash v_2 : \tau
    }{
        \Gamma; \Delta \vdash \texttt{(=/= \(v_1\) \(v_2\))} \goal
    }
    \and
    \infer{
        \exists \sigma \textrm{ where } \Delta \vdash \vec v : \sigma(\vec\tau)
    }{
        \Gamma, (R : \forall \vec\alpha . \vec\tau \rightarrow); \Delta \vdash \texttt{(\(R\) \(\vec v\))} \goal
    }
    \and
    \infer{ }{
        \Gamma; \Delta \vdash \texttt{(factor \(r\))} \goal
    }
\end{mathpar}
\vspace{-5pt}
\caption{Goal typing $\Gamma;\Delta \vdash g \goal$}
\label{fig:poly-goal-types}
\end{subfigure}
\smallskip

\begin{subfigure}{\textwidth}
\begin{mathpar}
    \infer{
        (R_1 : \forall \vec \alpha_1 . \vec \tau_1 \rightarrow), \dots, (R_N : \forall \vec \alpha_N . \vec \tau_N \rightarrow) ; \vec \alpha_1:*, \vec x_1:\vec \tau_1 \vdash g_1 \goal
        \\\\
        \vdots
        \\\\
        (R_1 : \forall \vec \alpha_1 . \vec \tau_1 \rightarrow), \dots, (R_N : \forall \vec \alpha_N . \vec \tau_N \rightarrow) ; \vec \alpha_N:*, \vec x_N:\vec \tau_N \vdash g_N \goal
    }{
        \texttt{(defrel (\(R_1\) \(\forall \vec \alpha_1\).(\(\vec x_1:\vec \tau_1\))) \(g_1\)))}, \dots,
        \texttt{(defrel (\(R_N\) \(\forall \vec \alpha_N\).(\(\vec x_N:\vec \tau_N\))) \(g_N\)))} \program
    }
\end{mathpar}
\vspace{-5pt}
\caption{Relation typing $\mathit{Rel},\dotsc \program$}
\label{fig:poly-relation-types}
\end{subfigure}
\smallskip

\begin{subfigure}{\textwidth}
\begin{mathpar}
  \infer{
    \Gamma;\vec{x}:\vec{\tau} \vdash g \goal
  }{
    \Gamma \vdash \texttt{(run (($\vec{x}:\vec{\tau}$)) $g$)} \query
  }
\end{mathpar}
\vspace{-5pt}
\caption{Query typing $\Gamma \vdash Q \query$}
\label{fig:poly-query-types}
\end{subfigure}

\caption{The type system of polymorphic semiringKanren}
\label{fig:poly-type}
\end{figure}

The type system of polymorphic semiringKanren is given in \cref{fig:poly-type}.
We abbreviate $\alpha_{1},\dots,\alpha_{n}$ as $\vec{\alpha}$, $\tau_{1},\dots,\tau_{n}$ as $\vec{\tau}$, $v_{1},\dots,v_{n}$ as $\vec{v}$, and $(x_{1}:\tau_{1}) \dots (x_{n}:\tau_{n})$ as $(\vec{x}:\vec{\tau})$.

The type validity judgment $\Delta \vdash \tau \type$ in \cref{fig:poly-types-valid} checks that each type variable~$\alpha$ used in type~$\tau$ is bound by $\alpha:*$ in the value typing context~$\Delta$.
In \cref{fig:poly-value-types,fig:poly-goal-types,fig:poly-relation-types,fig:poly-query-types}, we assume all types are valid and omit the premises for type validity.

The value typing judgment $\Delta \vdash v : \tau$ in \cref{fig:poly-value-types} checks that the value~$v$ is well-typed by consulting the types of logic variables $x:\tau$ in the value typing context~$\Delta$.
The goal typing judgment $\Gamma;\Delta \vdash g \goal$ in \cref{fig:poly-goal-types} uses two kinds of contexts:
value typing contexts and relation typing contexts ~$\Gamma$, which give the argument types of relations.
When a goal is a relation call, the arguments~$\vec v$ must inhabit the types $\vec \tau$ recorded in $\Gamma$ for the called relation, after some \emph{type variable substitution} $\sigma$ is applied to $\tau$.
In other words, we must have $\Delta \vdash \vec v : \sigma(\vec \tau)$.
We use $\sigma(\tau)$ to notate substituting concrete types for type variables in a type $\tau$, and $\sigma(\vec\tau)$ to notate substituting concrete types for all type variables in a list of types $\vec\tau$.
These substitutions are \emph{total}.
All type variables in $\tau$ are replaced with concrete types.

Relations and queries are checked in \cref{fig:poly-relation-types,fig:poly-query-types} respectively, using a relation typing context $\Gamma = (R_1 : \forall \vec \alpha_1 . \vec \tau_1 \rightarrow), \dots, (R_N : \forall \vec \alpha_N . \vec \tau_N \rightarrow)$ that lists all the relations $R_1,\dotsc,R_N$ in the program.

\subsection{Monomorphizing semantics}\label{section:semantics}

\begin{figure}
\begin{subfigure}{\linewidth}
\[
  \begin{aligned}
    &\textrm{Concrete values} & \varv
    &::= \texttt{sole} \quad|\quad
      \texttt{(pair $\varv$ $\varv$)} \quad|\quad
      \texttt{(left $\varv$)} \quad|\quad
      \texttt{(right $\varv$)} \\
    &\textrm{Value environments} & \delta &::= \varnothing \quad|\quad \delta,x \mapsto \varv\\
    &\textrm{Relation environments} & \gamma &::= \varnothing \quad|\quad \gamma,R_{\sigma} \mapsto \mathbb{W}
  \end{aligned}
\]
\caption{Concrete values and value environments}
\label{fig:poly-value-env}
\end{subfigure}

\begin{subfigure}{\linewidth}
\begin{align}
  \den{\texttt{Unit}}(\sigma)
  &= \{\texttt{sole}\} \\
  \den{\texttt{(Prod \(\tau_1\) \(\tau_2\))}}(\sigma)
  &= \{\texttt{(pair $\varv_{1}$ $\varv_{2}$)} \mid \varv_{1} \in \den{\tau_1}(\sigma),
    \varv_{2} \in \den{\tau_2}(\sigma)\}\\
  \den{\texttt{(Sum \(\tau_1\) \(\tau_2\))}}(\sigma)
  &= \{\texttt{(left $\varv$)} \mid \varv \in \den{\tau_1}(\sigma)\} \cup
    \{\texttt{(right $\varv$)} \mid \varv \in \den{\tau_2}(\sigma)\}\\
  \den{\alpha}(\sigma) &= \den{\sigma(\alpha)}
\end{align}
\caption{Denotational semantics for types}
\label{fig:poly-types-semantics}
\end{subfigure}

\begin{subfigure}{\linewidth}
\begin{align}
	\den{\texttt{sole}}(\delta) &= \texttt{sole} \\
	\den{\texttt{(pair $v_{1}$ $v_{2}$)}}(\delta) &= \texttt{(pair $\den{v_{1}}(\delta)$ $\den{v_{2}}(\delta)$)} \\
	\den{\texttt{(left $v$)}}(\delta) &= \texttt{(left $\den{v}(\delta)$)} \\
	\den{\texttt{(right $v$)}}(\delta) &= \texttt{(right $\den{v}(\delta)$)} \\
	\den{x} (\delta) &= \delta(x)
\end{align}
\caption{Denotational semantics for values}
\label{fig:poly-value-semantics}
\end{subfigure}
\caption{The denotational semantics of polymorphic semiringKanren}
\label{fig:poly-den}
\end{figure}

\begin{figure}
\begin{subfigure}{\linewidth}
  \begin{align}
    \den{\texttt{(conj \(g_1\) \(g_2\))}} (\gamma;\delta;\sigma)
    &= \den{g_1} (\gamma;\delta;\sigma) \times \den{g_2} (\gamma;\delta;\sigma) \\
	\den{\texttt{(disj \(g_1\) \(g_2\))}} (\gamma;\delta;\sigma)
    &= \den{g_1} (\gamma;\delta;\sigma) + \den{g_2} (\gamma;\delta;\sigma) \\
	\den{\texttt{(fresh ((\(x : \tau\))) \(g\))}} (\gamma;\delta;\sigma)
    &= \sum_{\varv \in \den{\sigma(\tau)}} \den{g} (\gamma;\delta,x \mapsto \varv;\sigma) \\
	\den{\texttt{(== \(v_1\) \(v_2\))}} (\gamma;\delta;\sigma)
    &=
      \begin{cases}
        1 & \textrm{if } \den{v_1} (\delta) = \den{v_2} (\delta) \\
        0 & \textrm{if } \den{v_1} (\delta) \ne \den{v_2} (\delta)
      \end{cases} \\
	\den{\texttt{(=/= \(v_1\) \(v_2\))}} (\gamma;\delta;\sigma)
    &=
      \begin{cases}
        1 & \textrm{if } \den{v_1} (\delta) \ne \den{v_2} (\delta) \\
        0 & \textrm{if } \den{v_1} (\delta) = \den{v_2} (\delta)
      \end{cases} \\
    \den{\texttt{(\(R\) \(\vec v\))}} (\gamma;\delta;\sigma)
    &=
    \gamma(R_{\sigma'})(\den{\vec v}(\delta))\\
      & \textrm{ where } \sigma' \textrm{ typechecks }
        \infer{\Delta \vdash \vec v : \sigma'(\vec \tau)}{\Gamma, (R : \vec \tau \rightarrow); \Delta \vdash (R \; \vec v) \goal}\nonumber
    \\
	\den{\texttt{(factor \(r\))}} (\gamma;\delta;\sigma) &= r
  \end{align}
\caption{Denotational semantics for goals}
\label{fig:poly-goals-semantics}
\end{subfigure}

\begin{subfigure}{\linewidth}
\begin{align}
    \den{\texttt{(defrel ($R$ $\forall \vec \alpha$.$(\vec x : \vec \tau)$) $g$)}}(\gamma)(\vec \varv)
    &=
    \dots, R_{\sigma_i} \mapsto \den{g}(\gamma; \vec x \mapsto \vec \varv; \sigma_i), \dots
    \\
	\den{\textit{Rel}_1, \dots, \textit{Rel}_Q}
    &=
	\; \mathrel{\textrm{fix}} \gamma \mathrel{\textrm{in}} \den{\textit{Rel}_1} (\gamma), \dots, \den{\textit{Rel}_Q} (\gamma)
\end{align}
\caption{Denotational semantics for relations}
\label{fig:poly-relations-semantics}
\end{subfigure}

\begin{subfigure}{\linewidth}
  \begin{align}
    \den{\texttt{(run (($\vec{x}:\vec\tau$)) $g$)}}(\gamma ; \vec\varv) 
    &= 
      \den{g}(\gamma ; \vec{x} \mapsto \vec{\varv}; \varnothing) 
  \end{align}
\caption{Denotational semantics for queries}
\label{fig:poly-query-semantics}
\end{subfigure}

\caption{The denotational semantics of polymorphic semiringKanren (continued)}
\label{fig:poly-den-cont}
\end{figure}

We present the denotational semantics for polymorphic semiringKanren in \cref{fig:poly-den,fig:poly-den-cont}.
\emph{Concrete values}, \emph{value environments}, and \emph{relation environments} are defined in \cref{fig:poly-value-env}.
Concrete values are values without logic variables.
Value environments are maps from logic variables to concrete values.
We use $\sigma(\Delta)$ to notate substituting concrete types for all type variables in the type environment $\Delta$,
and $\delta : \sigma(\Delta)$ to notate a value environment mapping logic variables in $\Delta$ to concrete values of the substituted types.
Relation environments are maps from relation variables (subscripted with a specific substitution) to weight arrays~$\mathbb{W}$, which are indexed by concrete values.

Each type denotes a finite set of concrete values, as shown in \cref{fig:poly-types-semantics}.
We handle type variables~$\alpha$ by applying a type variable substitution~$\sigma$ before taking a denotation.
Each value denotes a function from value environments~$\delta$ to concrete values, as shown in \cref{fig:poly-value-semantics}.

Each goal denotes a function that takes as arguments a relation environment~$\gamma$, a value environment~$\delta$, and a type variable substitution~$\sigma$ and returns a weight.
This is shown in \cref{fig:poly-goals-semantics}.
Goals can be thought of as arrays indexed by a value environment, and parameterized by a relation environment.

Different versions of each relation are generated for each type variable substitution in use.
We call these \emph{monomorphic instances} of polymorphic relations, and represent them by subscripting the polymorphic relation variable with the relevant substitution.
As shown in \cref{fig:poly-relations-semantics}, the denotation of a relation is a map of monomorphic instance names to monomorphic instance denotations,
and the denotation of a monomorphic instance is a weight array, as occurs in $\gamma$.
The denotation of a program is a fixed-point of $\gamma$, when iteratively used to take the denotation of all monomorphic instances of all relations.
Depending on the underlying semiring and the specific programs, this iterative process may not terminate.

\section{Equality Patterns and Large-Enough Relation Instances}

In this section, we explain our core contribution: an implementation strategy for polymorphic relations that does not require separately computing or storing every used instance of a polymorphic relation.
Instead, we can generalize \emph{large-enough} instances to larger used instances on the fly.
We first discuss an intuitive example, then describe our formalization.

\subsection{Revisiting Sum-Swap} \label{sec:revisiting-sum-swap}

Let us revisit \lstinline{sum-swap}.
Recall that the matrix structure of \lstinline{sum-swap} has identity matrices in the upper right and lower left corners.
When $|\alpha| = |\beta| = 3$, this ends up as follows:

\begin{equation}
\label{e:matrix6}
    \begin{bmatrix}
      0 & 0 & 0 & 1 & 0 & 0 \\
      0 & 0 & 0 & 0 & 1 & 0 \\
      0 & 0 & 0 & 0 & 0 & 1 \\
      1 & 0 & 0 & 0 & 0 & 0 \\
      0 & 1 & 0 & 0 & 0 & 0 \\
      0 & 0 & 1 & 0 & 0 & 0
    \end{bmatrix}
\end{equation}
Imagine we wish to instead call \lstinline{sum-swap} where $|\alpha| = 3$, but $|\beta| = 4$.
The resulting matrix will have $7$ rows and columns.
We can generate this matrix by ``looking up'' appropriate entries in the existing matrix where $|\alpha| = |\beta| = 3$.
First, observe that all entries in the existing matrix where \lstinline{x = (left ...)} and \lstinline{y = (left ...)} fail.
These correspond to the entries in the top left.
The elided parts of \lstinline{x} and \lstinline{y} are of types $\alpha$ and~$\beta$, and do not matter.
Similarly for entries in the bottom right, where both \lstinline{x} and \lstinline{y} are \lstinline{(right ...)}.
Thus, we can guess that all such entries should fail for our new matrix as well:
\begin{equation}
    \begin{bmatrix}
      0 & 0 & 0 & 0 & & & \\
      0 & 0 & 0 & 0 & & & \\
      0 & 0 & 0 & 0 & & & \\
        & & & & 0 & 0 & 0 \\
        & & & & 0 & 0 & 0 \\
        & & & & 0 & 0 & 0 \\
        & & & & 0 & 0 & 0
    \end{bmatrix}
\end{equation}
Next observe that when \lstinline{x = (left v)} and \lstinline{y = (right w)},
we have success when \lstinline{v = w} and failure otherwise.
Here, \lstinline{v} and \lstinline{w} have type $\alpha$, and all that matters is whether they are equal or not.
This gives us our identity matrix structure in the top right.
Similarly for type $\beta$ in the lower left.
\begin{equation}
\label{e:matrix7}
    \begin{bmatrix}
      0 & 0 & 0 & 0 & 1 & 0 & 0 \\
      0 & 0 & 0 & 0 & 0 & 1 & 0 \\
      0 & 0 & 0 & 0 & 0 & 0 & 1 \\
      1 & 0 & 0 & 0 & 0 & 0 & 0 \\
      0 & 1 & 0 & 0 & 0 & 0 & 0 \\
      0 & 0 & 1 & 0 & 0 & 0 & 0 \\
      0 & 0 & 0 & 1 & 0 & 0 & 0
    \end{bmatrix}
\end{equation}

This example demonstrates a general implementation strategy for polymorphic relations:
When generalizing a smaller matrix to a larger matrix, for each entry in the larger matrix,
we look up an entry in the smaller matrix with the same \emph{equality pattern}.

\subsection{Shells and Holes}

Before defining equality patterns, we first need notions of \emph{shells} and \emph{holes}.
Recall that our denotational semantics evaluates a goal at a value environment~$\delta$, which inhabits a value typing context~$\Delta$.
When working with polymorphic relations, we use type variable substitutions~$\sigma$ to convert variable types into concrete ones.
For example, the top-level typing context of \texttt{sum-swap} is $\Delta = \alpha : *, \beta : *, x : \texttt{(Sum $\alpha$ $\beta$)}, y : \texttt{(Sum $\beta$ $\alpha$)}$.
The specific version of \lstinline{sum-swap} where $|\alpha|=|\beta|=3$ might be evaluated at some value environment $\delta : \sigma(\Delta)$ where $\sigma(\alpha) = \sigma(\beta) = \texttt{(Sum Unit (Sum Unit Unit))}$.

Suppose $\varnothing \vdash v : \sigma(\tau)$.
Note that $v$ here cannot contain logic variables, and inhabits a concrete type.
The \emph{shell} of $v$ under $\tau$ is the parts of $v$ that are not variable-typed in $\tau$ pre-substitution.
We define $\shell_\tau(v)$ below.
If we do reach a part of $v$ that has variable type $\alpha$ under $\tau$, we call it \lstinline[mathescape]{(hole $\alpha$)}.
\begin{align}
  \shell_\texttt{Unit}(\texttt{sole}) &= \texttt{sole} \\
  \shell_\texttt{(Prod \(\tau_1\) \(\tau_2\))}(\texttt{(pair \(v_1\) \(v_2\))}) &= \texttt{(pair \(\shell_{\tau_1}(v_1)\) \(\shell_{\tau_2}(v_2)\))} \\
  \shell_\texttt{(Sum \(\tau_1\) \(\tau_2\))}(\texttt{(left \(v\))}) &= \texttt{(left \(\shell_{\tau_1}(v)\))} \\
  \shell_\texttt{(Sum \(\tau_1\) \(\tau_2\))}(\texttt{(right \(v\))}) &= \texttt{(right \(\shell_{\tau_2}(v)\))} \\
  \shell_\alpha(v) &= \texttt{(hole \(\alpha\))}
\end{align}

For each type variable $\alpha$, we can also get a list of the parts of $v$ which would be variable-typed pre-substitution.
These are the $\alpha$-holes of $v$ under $\tau$.
We define $\holes_{\alpha \in \tau}(v)$ below.
\begin{align}
  \holes_{\alpha \in \texttt{Unit}}(\texttt{sole}) &= \varnothing \\
  \holes_{\alpha \in \texttt{(Prod \(\tau_1\) \(\tau_2\))}}(\texttt{(pair \(v_1\) \(v_2\))}) &= \holes_{\alpha \in \tau_1}(v_1), \holes_{\alpha \in \tau_2}(v_2) \\
  \holes_{\alpha \in \texttt{(Sum \(\tau_1\) \(\tau_2\))}}(\texttt{(left \(v\))}) &= \holes_{\alpha \in \tau_1}(v) \\
  \holes_{\alpha \in \texttt{(Sum \(\tau_1\) \(\tau_2\))}}(\texttt{(right \(v\))}) &= \holes_{\alpha \in \tau_2}(v) \\
  \holes_{\alpha \in \alpha}(v) &= v \\
  \holes_{\beta \in \alpha}(v) &= \varnothing \qquad\text{if $\beta\ne\alpha$}
\end{align}

Given a shell and the $\alpha$-holes for all $\alpha$, we can plug in holes left-to-right into the appropriately-typed \lstinline[mathescape]{(hole $\dots$)} parts of the shell to get back $v$.

As an example, consider
$\texttt{(pair (left sole) (right sole))} : \sigma(\texttt{(Prod $\alpha$ $\alpha$)})$
where $\sigma(\alpha)=\texttt{(Sum Unit Unit)}$.
We have
\begin{align}
\shell_{\texttt{(Prod $\alpha$ $\alpha$)}}(\texttt{(pair (left sole) (right sole))}) &= \texttt{(pair (hole $\alpha$) (hole $\alpha$))}\\
\holes_{\alpha \in \texttt{(Prod $\alpha$ $\alpha$)}}(\texttt{(pair (left sole) (right sole))}) &= \texttt{(left sole)}, \texttt{(right sole)}
\end{align}

Given a value environment $\delta : \sigma(\Delta)$, the shell of $\delta$ is a map from logic variables $x$ to $\shell_{\tau}(\delta(x))$ for every $x : \sigma(\tau)$ in $\Delta$, and the hole of $\delta$ under a type variable $\alpha$ in $\Delta$ is a map from logic variables $x$ to $\holes_{\alpha \in \tau}(\delta(x))$.

\subsection{Equality Patterns}
\label{sec:equality-patterns}

Consider value environments $\delta_1$ and $\delta_2$ where $\delta_1 : \sigma_1(\Delta)$ and $\delta_2 : \sigma_2(\Delta)$.
We say $\delta_1$ and $\delta_2$ have the same \emph{equality pattern} under $\Delta$, notated $\delta_1 \eqpatD \delta_2$,
when their shells are equal,
and each pair of holes in $\delta_1$ is equal if and only if it is also equal in $\delta_2$.
In other words, pairs of holes that are equal in $\delta_1$ must also be equal in $\delta_2$, and pairs of holes that are disequal in $\delta_1$ must also be disequal in $\delta_2$.
Note that the the second condition only makes sense if the former holds: $\delta_1$ and $\delta_2$ only have the same holes if their shells are equal.

For example, consider \(\delta_1 : \sigma_1(\Delta)\) and \(\delta_2 : \sigma_2(\Delta)\), with \(\Delta\) as follows:
\begin{equation}
	\Delta = \alpha: *, x : \texttt{(Sum \(\alpha\) \(\alpha\))}, y : \alpha
\end{equation}
We will present potential equality patterns of the form \(\delta_1 \eqpatD \delta_2\), and explain why they do or do not hold.
We underline holes for illustration.
First consider:
\begin{equation}
	x \mapsto \texttt{(left \underline{sole})}, y \mapsto \texttt{\underline{sole}} \eqpatD x \mapsto \texttt{(right \underline{sole})}, y \mapsto \texttt{\underline{sole}} 
\end{equation}
Here \(\sigma_1(\alpha) = \sigma_2(\alpha) = \texttt{Unit}\).
This equality pattern does \emph{not} hold, because the shells of \(x\), \(\texttt{(left \(\dots\))}\) and \(\texttt{(right \(\dots\))}\), are not equal.
\begin{multline}
	x \mapsto \texttt{(left \(\underline{\texttt{(right sole)}}\))}, y \mapsto \underline{\texttt{(left sole)}} \\\eqpatD x \mapsto \texttt{(left \(\underline{\texttt{(right sole)}}\))}, y \mapsto \underline{\texttt{(right sole)}} 
\end{multline}
Here \(\sigma_1(\alpha) = \sigma_2(\alpha) = \texttt{(Sum Unit Unit)}\).
This equality pattern does \emph{not} hold, because the holes are disequal under \(\delta_1\) (\(\texttt{(right sole)} \ne \texttt{(left sole)}\)), but equal under \(\delta_2\) (\(\texttt{(right sole)} = \texttt{(right sole)}\)).
\begin{multline}
	x \mapsto \texttt{(left \(\underline{\texttt{(right sole)}}\))}, y \mapsto \underline{\texttt{(left sole)}} \\\eqpatD x \mapsto \texttt{(left \(\underline{\texttt{(left sole)}}\))}, y \mapsto \underline{\texttt{(right sole)}} 
\end{multline}
Here again \(\sigma_1(\alpha) = \sigma_2(\alpha) = \texttt{(Sum Unit Unit)}\).
This equality pattern \emph{does} hold. The shells are equal,
and the holes are disequal both under \(\delta_1\) (\(\texttt{(right sole)} \ne \texttt{(left sole)}\)) and under \(\delta_2\) (\(\texttt{(left sole)} \ne \texttt{(right sole)}\)).
\begin{equation}
	x \mapsto \texttt{(left \(\underline{\texttt{(left sole)}}\))}, y \mapsto \underline{\texttt{(left sole)}} \eqpatD x \mapsto \texttt{(left \underline{sole})}, y \mapsto \texttt{\underline{sole}} 
\end{equation}
Here \(\sigma_1(\alpha) = \texttt{(Sum Unit Unit)}\) and \(\sigma_2(\alpha) = \texttt{Unit}\).
This equality pattern \emph{does} hold. The shells are equal,
and the holes are equal both under \(\delta_1\) (\(\texttt{(left sole)} = \texttt{(left sole)}\)) and under \(\delta_2\) (\(\texttt{sole} = \texttt{sole}\)).

\subsection{Large-Enough Relation Instances}

For a polymorphic relation \(R : \forall \vec\alpha . \vec\tau \rightarrow\),
we say that one monomorphic instance~\(R_{\sigma_2}\) is \emph{larger} than another instance~\(R_{\sigma_1}\)
if \(|\sigma_2(\alpha)| \ge |\sigma_1(\alpha)|\) for all \(\alpha \in \vec \tau\).
The inequality compares the sizes of two concrete types.
Intuitively, we say that the particular instance~\(R_{\sigma_1}\) is \emph{large-enough}
when we can recreate the weight array representing the larger instance \(R_{\sigma_2}\) from the smaller instance~\(R_{\sigma_1}\).

We have seen that \texttt{two-valued} behaves differently when called with \texttt{Unit} and \texttt{(Sum Unit Unit)}.
The \texttt{Unit}-typed monomorphic instance of \texttt{two-valued} is not large-enough, but the \texttt{(Sum Unit Unit)} instance is; we expect \texttt{two-valued} to succeed when its type contains two or more values.

To expand a smaller matrix like \eqref{e:matrix6} to a larger matrix like \eqref{e:matrix7}, recall that our approach is to look up indices with similar equality patterns.
In the most extreme case, we might want to look up an index in the smaller matrix where all holes are disequal.
To ensure that such an index exists, we need the type to be inhabited by as many values as there are holes:
if~a value environment can contain $n$ holes of type~\(\alpha\), then we need to instantiate \(\alpha\) with a type whose size is at least~\(n\), so that it can be inhabited by \(n\) disequal values for the equality pattern to hold.
Holes in value environments correspond to type variables in value typing contexts.
So to figure out how large an instance needs to be in order to be large-enough, we can count the number of times a type variable occurs in the relation's value typing contexts.

Given a derivation of a goal typing judgment \(\Gamma;\Delta \vdash g \goal\), let \(\#_\alpha g(\Delta)\) denote the maximum number of times the type variable~\(\alpha\) occurs in \emph{any} value typing context in the derivation.
We can calculate \(\#_\alpha g(\Delta)\) by traversing the syntax tree of $g$, extending $\Delta$ when we encounter fresh variables, and counting the maximum number of times $\alpha$-typed values can occur in the variables in $\Delta$.
For example, given $\Delta = \alpha:*, x : \texttt{(Sum \(\alpha\) \(\alpha\))}, y : \texttt{(Prod \(\alpha\) \(\alpha\))}$,
$\alpha$ occurs three times in $\Delta$: once in $x$ (because either \texttt{left} or \texttt{right} holds $\alpha$ only once), and twice in $y$.

This count provides a lower bound for type variable sizes for a relation to be large-enough.
In practice, if the definition of a relation includes calls to other relations, these calls may impose larger type variable size requirements.


Formally, given a relation environment~\(\gamma : \Gamma\), we say an instance~\(R_{\sigma_1}\) of a relation \(R : \forall \vec\alpha . \vec\tau \rightarrow\) is \emph{large-enough} if
\begin{enumerate}
\item for all larger instances \(R_{\sigma_2}\) of~\(R\),
\item for all value typing contexts \(\Delta\) that bind exactly the type variables~\(\vec\alpha\),
\item for all arguments \(\vec v\) such that \(\Delta \vdash \vec v : \vec \tau\),
\item and for all value environments \(\delta_1 : \sigma_1(\Delta)\) and \(\delta_2 : \sigma_2(\Delta)\) such that \(\delta_1 \eqpatD \delta_2\),
\end{enumerate}
we have
\(
    \gamma(R_{\sigma_1})(\den{\vec v}(\delta_1))
=
    \gamma(R_{\sigma_2})(\den{\vec v}(\delta_2))
\).

To illustrate this definition, we briefly reinvestigate \lstinline{equal} in more detail.
The denotations of monomorphic instances of \lstinline{equal} are as follows, for increasing sizes of $\alpha$:
\begin{equation}
    \begin{bmatrix}
        1
    \end{bmatrix}
    ,
    \begin{bmatrix}
        1 & 0 \\
        0 & 1
    \end{bmatrix}
    ,
    \begin{bmatrix}
        1 & 0 & 0 \\
        0 & 1 & 0 \\
        0 & 0 & 1
    \end{bmatrix}
    ,
    \begin{bmatrix}
        1 & 0 & 0 & 0 \\
        0 & 1 & 0 & 0 \\
        0 & 0 & 1 & 0 \\
        0 & 0 & 0 & 1
    \end{bmatrix}
    ,
    \dots
\end{equation}
We aim to show that the two-by-two identity matrix is large-enough.
For $\vec v$ to typecheck under $\Delta$, it must have logic variables at the top-level
(because only logic variables can inhabit type variables).
So, we can disregard shells and directly check that the weights match for specific value environments with the same equality pattern for holes.
For any matrix larger than two-by-two, we can see that the weight $0$ occurs when the two logic variables in the environment are disequal (on the matrix off-diagonals).
This is matched in the two-by-two matrix.
Similarly, the weight $1$ occurs when the two logic variables are equal (on the matrix diagonals).
This is also matched by the two-by-two matrix.
Thus, the two-by-two matrix is large-enough.

\subsection{Equality Patterns Preserve Goal Weight}
\label{sec:eqpat-preserves-weight}

As progress towards our compilation technique, we note that the denotation weight of a goal is consistent under value environments with the same equality pattern.
\begin{theorem}
	\label{thm:eqpat-weight}
	Let \(\Gamma; \Delta \vdash g \goal\), \(\delta_1 : \sigma_1(\Delta)\), \(\delta_2 : \sigma_2(\Delta)\),
	and all relations called within \(g\)\ be large-enough.
	If 
    \begin{enumerate}
        \item \(\delta_1 \eqpatD \delta_2\),
        \item for any \(\alpha \in \Delta\),  \(|\sigma_1(\alpha)| \ge \#_\alpha g(\Delta)\) and \(|\sigma_2(\alpha)| \ge \#_\alpha g(\Delta)\),
        \item the \(+\) operation in the weight semiring is idempotent, i.e., \(r + r = r\) for all~\(r\),
    \end{enumerate}
    then
	\(\den{g}(\gamma; \delta_1; \sigma_1) = \den{g}(\gamma; \delta_2; \sigma_2)\).
\end{theorem}
A full proof of this theorem can be found in \cite{volkov2026}.
The proof follows from induction on the goal structure of $g$.
The \lstinline{conj} and \lstinline{disj} cases follow from straightforward application of the inductive hypothesis, and the \lstinline{==} and \lstinline{=/=} cases follow from the definition of equality patterns.
The relation call case holds because we assume all relations are large-enough; this lets us delay dealing with the type variable size constraints imposed by relation calls.
The \lstinline{factor} case is trivial.

We focus now on the \lstinline{fresh} case.
The third theorem requirement, that semiring addition be idempotent, is crucial both here and later;
in particular, it allows \lstinline{fresh} to behave consistently even when using type variables that have different sizes after substitution.
Let $g'$ be some subgoal.
We wish to show:
\begin{equation}
\label{e:fresh-case}
    \sum_{i \in \sigma_1(\tau)} \den{g'}(\gamma;\delta_1,x \mapsto i;\sigma_1)
    = 
    \sum_{j \in \sigma_2(\tau)} \den{g'}(\gamma;\delta_2,x \mapsto j;\sigma_2)
\end{equation}
We can show 
that for each $i$, we can find a $j$ such that $\delta_1, x \mapsto i \eqpat_{\Delta,\tau} \delta_2, x \mapsto j$,
and vice versa.
From this and by applying the inductive hypothesis on $\den{g'}(\gamma;\delta_1,x \mapsto i;\sigma_1) \eqpat_{\Delta,\tau} \den{g'}(\gamma;\delta_2,x \mapsto j;\sigma_2)$,
we can deduce that that each weight that appears on the left-hand side of \eqref{e:fresh-case} must appear at least once on the right-hand side,
and vice versa.
In other words, the two sides of \eqref{e:fresh-case} sum over the same set of weights, just repeated possibly different numbers of times---for example, one side might be $r_1+r_2+r_2+r_1$ while the other side might be $r_2+r_1+r_2$.
But because addition is idempotent, we can collapse groups of equal weights to conclude that the two sides are equal.

\section{Compilation}

\subsection{Compiling Polymorphic Relation Calls}

We can mostly avoid the need for monomorphization by only evaluating the smallest large-enough instances of polymorphic relations,
and wrapping calls at larger types using the expansion process described in \cref{sec:revisiting-sum-swap}.
We do not have a language primitive that allows us to do this directly,
but can achieve the same result via a method of \emph{enforcing equality patterns}:
we first call the smallest large-enough relation instance with smaller-typed fresh variables,
and enforce an equality pattern between those fresh variables and the larger-typed relation call arguments.

\Cref{sec:equality-patterns} defined equality patterns for environments.
Here, it makes sense to think of equality patterns between sub-environments.
One sub-environment is for the smaller-typed fresh variables, and the other is for the larger-typed relation call arguments.

We can generate as follows the semiringKanren code which succeeds when two sub-environments have the same equality pattern:
\begin{equation}
\begin{aligned}
	& \enforceeqpatD(\Delta_1; \Delta_2) = \\
	& \quad \texttt{(fresh (\textrm{ancillary variables for \(\envshell\) and \(\envholes\)})} \\
	& \quad \quad \texttt{(conj }\\
	& \quad \quad \quad \textrm{deconstruct \(\Delta_1\) and \(\Delta_2\) into the ancillary variables} \\
	& \quad \quad \quad \texttt{(== \(\envshell_\Delta(\Delta_1)\) \(\envshell_{\Delta}(\Delta_2)\))} \\
	& \quad \quad \quad \textrm{for \(i,j \in 1,\dots,\#_\alpha\Delta\)} \\
	& \quad \quad \quad \quad \texttt{(disj} \\
	& \quad \quad \quad \quad \quad \texttt{(conj} \\
	& \quad \quad \quad \quad \quad \quad \texttt{(== \(\envholes_{\alpha \in \Delta}(\Delta_1)[i]\) \(\envholes_{\alpha \in \Delta}(\Delta_1)[j]\))} \\
	& \quad \quad \quad \quad \quad \quad \texttt{(== \(\envholes_{\alpha \in \Delta}(\Delta_2)[i]\) \(\envholes_{\alpha \in \Delta}(\Delta_2)[j]\)))} \\
	& \quad \quad \quad \quad \quad \texttt{(conj} \\
	& \quad \quad \quad \quad \quad \quad \texttt{(=/= \(\envholes_{\alpha \in \Delta}(\Delta_1)[i]\) \(\envholes_{\alpha \in \Delta}(\Delta_1)[j]\))} \\
	& \quad \quad \quad \quad \quad \quad \texttt{(=/= \(\envholes_{\alpha \in \Delta}(\Delta_2)[i]\) \(\envholes_{\alpha \in \Delta}(\Delta_2)[j]\))))} \\
	& \quad \quad \quad \dots \texttt{))}
\end{aligned}
\end{equation}
The generated code needs to be consistent regardless of the specific values in the sub-environments,
so we express it in terms of type sub-environments (which include variable names).
We use $\envshell$ and $\envholes$ as ``vectorized'' versions of $\shell$ and $\holes$, which act on the each of the current values inhabited by the passed sub-environments.
The generated code closely matches the definition of equality patterns.
We first deconstruct the environments into shells and holes (using the fresh ancillary variables), and check that the shells are equal.
Then for each pair of holes, we check that they are either equal under both sub-environments or disequal under both sub-environments.
While these operations are defined in terms of $\Delta_1$ and $\Delta_2$, in practice they operate on the ancillary variables.

For example, consider enforcing an equality pattern between the sub-environment
\[\Delta_1 = x_1 : \texttt{(Prod Unit (Num 2))}, y_1 : \texttt{(Prod Unit (Num 2))}\]
and the sub-environment
\[\Delta_2 = x_2 : \texttt{(Prod Unit (Num 3))}, y_2 : \texttt{(Prod Unit (Num 3))},\]
both of which are based on the polymorphic sub-environment
\[\Delta = x : \texttt{(Prod Unit \(\alpha\))},y : \texttt{(Prod Unit \(\alpha\))}.\]
Here, we use \(\texttt{(Num \(n\))}\) as a shorthand for a sum type of size \(n\).
Calling $\enforceeqpatD$ yields the following expression:
\begin{equation}
\begin{aligned}
    & \enforceeqpatD(\Delta_1; \Delta_2) = \\
    & \quad \texttt{(fresh (x1hole : (Num 2)) (y1hole : (Num 2))} \\
    & \quad\quad\quad\quad\;\; \texttt{(x2hole : (Num 3)) (y2hole : (Num 3))} \\
    & \quad\quad \texttt{(conj} \\
    & \quad\quad\quad \textit{; check that shells have the same form, extracting holes where possible} \\
    & \quad\quad\quad \textit{; only one such shell is possible: \(x = \texttt{(pair sole \_)}, y = \texttt{(pair sole \_)}\)} \\
    & \quad\quad\quad \texttt{(== $x_1$ (pair sole x1hole)) (== $x_2$ (pair sole x2hole))} \\
    & \quad\quad\quad \texttt{(== $y_1$ (pair sole y1hole)) (== $y_2$ (pair sole y2hole))} \\
    & \quad\quad\quad \textit{; enforce an equality pattern between the holes of \(\Delta_1\) and \(\Delta_2\)} \\
    & \quad\quad\quad \texttt{(disj} \\
    & \quad\quad\quad\quad \textit{; either $\texttt{x1hole} = \texttt{y1hole}$ and $\texttt{x2hole} = \texttt{y2hole}$} \\
    & \quad\quad\quad\quad \texttt{(conj (== x1hole y1hole) (== x2hole y2hole))} \\
    & \quad\quad\quad\quad \textit{; or $\texttt{x1hole} \not{=} \texttt{y1hole}$ and $\texttt{x2hole} \not{=} \texttt{y2hole}$} \\
    & \quad\quad\quad\quad \texttt{(conj (=/= x1hole y1hole) (=/= x2hole y2hole)))))} \\
\end{aligned}
\end{equation}

From here, compiling relation calls is easy.
We generate fresh variables at the smaller types, replace the relation call, and enforce an equality pattern between the original larger argument environment and the fresh smaller argument environment.
Assuming $R_{\sigma_1}$ is our larger relation instance, $R_{\sigma_2}$ is the smallest large-enough relation instance, and $\Delta_1, \Delta_2$ are the larger and smaller environments respectively, we can use the following:
\begin{equation}
\begin{aligned}
	& \compile(\texttt{(\(R_{\sigma_1}\) \(\vec v\))}) = \\
	& \quad \texttt{(fresh (($\delta_1 : \Delta_1 = \sigma_1(\Delta)$) (\(\delta_2 : \Delta_2 = \sigma_2(\Delta)\))) } \\
	& \quad\quad \texttt{(conj} \\
    & \quad\quad\quad \texttt{(== $\delta_1$ $\vec v$)} \\
	& \quad\quad\quad \texttt{(\(R_{\sigma_2}\) \(\delta_2\))} \\
	& \quad\quad\quad \texttt{\(\enforceeqpatD(\Delta_1; \Delta_2)\)))} \\
\end{aligned}
\end{equation}

We assert that this compilation technique works correctly.
\begin{theorem}
	\label{thm:compile-poly}
	Given any \(\gamma : \Gamma\) and \(\sigma\), large-enough relation instance \(R_{\sigma_1}\), value environment \(\delta_1 : \Delta_1 =  \sigma_1(\Delta)\),
	and \(\vec v\) satisfying \(\Gamma; \Delta \vdash \texttt{(\(R\) \(\vec v\))} \goal\), it follows that
	\[ \den{\texttt{(\(R_{\sigma_1}\) \(\vec v\))}}(\gamma;\delta_1;\sigma) = \den{\compile(\texttt{(\(R_{\sigma_1}\) \(\vec v\))})}(\gamma;\delta_1;\sigma) \]
\end{theorem}
We have already argued in \cref{thm:eqpat-weight} that under the right conditions, calling relations with environments with the same equality pattern preserves the weight of the call.
The code generated by $\enforceeqpatD$ enforces an equality pattern between the original arguments and the smaller arguments used for the relation call,
so by \cref{thm:eqpat-weight}, the weight of the relation call is correct.
The outer fresh is an idempotent summation, so when $\enforceeqpatD$ fails or the relation call has multiple succeeding assignments, the overall weight does not change.

As a brief example, we aim to compile the following call to the polymorphic \lstinline{equal} relation:
\begin{lstlisting}
(equal (pair (left sole) (right sole)) (pair (left sole) (right sole)))
\end{lstlisting}
Here, \lstinline[mathescape]{$\alpha$ = (Prod (Sum Unit Unit) (Sum Unit Unit))}.
The smallest large-enough instance of \lstinline{equal} has $|\alpha| = 2$,
so we compile using \lstinline[mathescape]{$\alpha$ = (Sum Unit Unit)}.
The result is as follows:
\begin{lstlisting}[mathescape]
(fresh ((x1 : (Prod (Sum Unit Unit) (Sum Unit Unit)))
        (y1 : (Prod (Sum Unit Unit) (Sum Unit Unit)))
        (x2 : (Sum Unit Unit)) (y2 : (Sum Unit Unit)))
    (conj
        (== x1 (pair (left sole) (right sole)))
        (== y1 (pair (left sole) (right sole)))
        (equal$_{\alpha \mapsto \texttt{(Sum Unit Unit)}}$ x2 y2)
        $\enforceeqpatD($
          x1:(Prod (Sum Unit Unit) (Sum Unit Unit)),
          y1:(Prod (Sum Unit Unit) (Sum Unit Unit));
          x2:(Sum Unit Unit), y2:(Sum Unit Unit)$)$))
\end{lstlisting}

\subsection{Compiling Programs}

The condition that relation calls be large-enough is important.
We may still need to generate monomorphic instances for relation calls that are not large-enough.
The conditions for \cref{thm:eqpat-weight} give us a way of deciding whether relation instances are large-enough,
by examining their goal structures.
In particular, we need to check that the type variable sizes are larger than the number of type variable occurrences,
and that relation calls within the goal structure are large-enough.
Checking if relation calls are large-enough is difficult, as we aim to support recursive relations.

We propose a fixpoint process for checking whether relation calls are large-enough, and for determining the smallest large-enough relation instances.
For each relation, we track the minimum large-enough size for each type variable.
We start by assuming the minimum large-enough sizes are all $0$, and iteratively update this hypothesis.
For each type variable $\alpha$ of each relation, we traverse the goal structure, tracking the current value typing context $\Delta$.
When we reach a leaf node (\lstinline{==}, \lstinline{=/=}, or \lstinline{factor}),
we count the maximum number of times $\alpha$-typed values can occur in the current environment.
For \lstinline{conj} and \lstinline{disj}, we take the maximum count from either subgoals.
For \lstinline{fresh}, we extend the environment before taking the appropriate count from the subgoal.
For relation calls, we look up the previously-calculated minimum large-enough size for the type variables in that relation,
use that to calculate the minimum argument sizes,
and work backwards to find the needed type variable sizes within the current relation.
We then take the maximum of this value and the type variable count from the current environment.

Once we have determined the minimum large-enough sizes, and which calls are large-enough,
we generate the smallest large-enough instances of polymorphic relations
(along with any needed monomorphic instances for relation calls that are not large-enough).
We then compile large-enough relation calls as specified above to use these instances.
The result is a non-polymorphic semiringKanren program that can be evaluated as usual.

\section{Related Work}

\subsection{miniKanren}

The semiringKanren languge draws extensive inspiration from miniKanren~\cite{byrd2012}.
Particularly, semiringKanren's syntax is largely borrowed from the the microKanren~\cite{hemann2023} variant.
The miniKanren language is untyped and uses a top-down search strategy, supporting polymorphism by default.

\subsection{Weighted Bottom-Up Relational Languages}

Datalog~\cite{green2007} is the prototypical weighted bottom-up relational language.
Datalog program evalaution is based on finding the \emph{least fixed point} of a set of facts.
Datalog programs can be annotated with semiring weights.
Flix~\cite{madsen2020} is a descendent of datalog that supports evaluation over arbitrary semirings.
The evaluation strategy of semiringKanren is directly drawn from that of datalog.

Dyna~\cite{francislandau2024} is a descendant of datalog supporting \emph{aggregators}, which permit more operations on weights than plain semirings.
Currently, Dyna is implemented by applying a system of algebraic rules.

\subsection{Polymorphic Bottom-Up Relational Languages}

Datalog supports a form of ad-hoc polymorphism by default, where unary relations are used as types.
PolyDatalog~\cite{atzeni2010} extends this to a more sophisticated object-relational type system.

Existing bottom-up relational languages supporting parametric polymorphism include Flix~\cite{madsen2020} and functional IncA~\cite{pacak2022}.
Both of these languages are extensions of Datalog, and implement polymorphism via monomorphization.

Mercury~\cite{somogyi1996} is a top-down relational language which supports polymorphic relations
by passing specialized unification and comparison operations for each type variable.
A backend for Mercury has been implemented atop Aditi~\cite{vaghanl1994}, a bottom-up deductive database system,
but it is unclear if polymorphism for this backend is implemented using a similar method, or with monomorphization.

\subsection{Other Related Work}

Pixel arrays~\cite{spivak2017} approximate solutions to nonlinear systems of equations
by combining discretized plots using a generalized form of matrix multiplication.
Pixel arrays can be implemented in semiringKanren by representing the discretized equations as relations,
creating fresh variables for the hidden variables in the equations, and \lstinline{conj}ing calls to the relations.
This is mathematically equivalent to the generalized form of matrix multiplication.

The paper ``Testing Polymorphic Properties'' \cite{bernardy2010} outlines a method to test polymorphic functions in languages with parametric polymorphism
by only testing a single monomorphic instance of the function.
While our work is different, we also show that only a single monomorphic instance is needed when using a polymorphic relation.

\section{Future Work}

Prior work \cite{volkov2026}
demonstrates that higher-order relations can be encoded into product types if used affinely (zero or once).
We believe it may be possible to generalize this to full higher-order relations by adding a construct capable of copying to the language.

As of writing, semiringKanren does not support recursive types.
This is very limiting.
For example, semiringKanren cannot represent arbitrary lambda calculus terms and cannot perform the traditional quine generation of miniKanren-family languages.
Intuitively, implementing recursive types seems to be at odds with semiringKanren's denotation of finite arrays.
We see three potential ways of implementing recursive types in semiringKanren.
First, we could use \emph{gas-recursive types}, where recursive types are equipped with a size parameter, to establish an upper bound on the needed array sizes.
Second, it may be possible to generalize the approach presented by \citet{chiang2023}, which uses de/refunctionalization to eliminate recursive data types, to ``de/rerelationalization.''
Finally, if semiringKanren gains support for higher-order relations, it may be possible to use Church-style encodings of recursive types.

\section{Conclusion}

We present a method to compile polymorphic semiringKanren programs into non-polymorphic ones without monomorphization,
by using equality patterns and large-enough relation instances.
We give a monomorphizing semantics for polymorphic semiringKanren and demonstrated that the compilation technique preserves equivalence with this semantics.
We identify when this technique can and cannot be applied, and discuss practical implementation considerations.

We feel that the miniKanren approach to relational programming lends well to these developments.
We hope our method may find application in relational languages beyond semiringKanren.
In particular, we believe it may offer efficiency gains by reducing the need to recalculate the same relation at different types.
We also believe it can help enable ``separate evaluation'' of programs, so relations can be precomputed and reused without reevaluation.

\bibliographystyle{plainnat}
\bibliography{main}

@article{byrd2012,
    author = {Byrd, William and Holk, Eric and Friedman, Daniel},
    year = {2012},
    month = {09},
    pages = {},
    title = {miniKanren, live and untagged: quine generation via relational interpreters (programming pearl)},
    isbn = {978-1-4503-1895-2},
    doi = {10.1145/2661103.2661105}
}

@InProceedings{hemann2023,
  author="Hemann, Jason and Friedman, Daniel P.",
  editor="Chang, Stephen",
  title="Nearly Macro-free microKanren",
  booktitle="Trends in Functional Programming",
  year="2023",
  publisher="Springer Nature Switzerland",
  address="Cham",
  pages="72--91",
  isbn="978-3-031-38938-2"
}

@inproceedings{green2007,
    author = {Green, Todd J. and Karvounarakis, Grigoris and Tannen, Val},
    title = {Provenance semirings},
    year = {2007},
    isbn = {9781595936851},
    publisher = {Association for Computing Machinery},
    address = {New York, NY, USA},
    url = {https://doi.org/10.1145/1265530.1265535},
    doi = {10.1145/1265530.1265535},
    booktitle = {Proceedings of the Twenty-Sixth ACM SIGMOD-SIGACT-SIGART Symposium on Principles of Database Systems},
    pages = {31–40},
    numpages = {10},
    location = {Beijing, China},
    series = {PODS '07}
}

@misc{spivak2017,
      title={Pixel Arrays: A fast and elementary method for solving nonlinear systems},
      author={David I. Spivak and Magdalen R. C. Dobson and Sapna Kumari and Lawrence Wu},
      year={2017},
      eprint={1609.00061},
      archivePrefix={arXiv},
      primaryClass={math.NA},
      url={https://arxiv.org/abs/1609.00061},
}

@article{madsen2020,
    author = {Madsen, Magnus and Lhot\'{a}k, Ond\v{r}ej},
    title = {Fixpoints for the masses: programming with first-class Datalog constraints},
    year = {2020},
    issue_date = {November 2020},
    publisher = {Association for Computing Machinery},
    address = {New York, NY, USA},
    volume = {4},
    number = {OOPSLA},
    url = {https://doi.org/10.1145/3428193},
    doi = {10.1145/3428193},
    journal = {Proc. ACM Program. Lang.},
    month = nov,
    articleno = {125},
    numpages = {28}
}

@InProceedings{pacak2022,
    author =  {Pacak, Andr\'{e} and Erdweg, Sebastian},
    title =   {{Functional Programming with Datalog}},
    booktitle =   {36th European Conference on Object-Oriented Programming (ECOOP 2022)},
    pages =   {7:1--7:28},
    series =  {Leibniz International Proceedings in Informatics (LIPIcs)},
    ISBN =    {978-3-95977-225-9},
    ISSN =    {1868-8969},
    year =    {2022},
    volume =  {222},
    editor =  {Ali, Karim and Vitek, Jan},
    publisher =   {Schloss Dagstuhl -- Leibniz-Zentrum f{\"u}r Informatik},
    address = {Dagstuhl, Germany},
    URL =     {https://drops.dagstuhl.de/entities/document/10.4230/LIPIcs.ECOOP.2022.7},
    URN =     {urn:nbn:de:0030-drops-162354},
    doi =     {10.4230/LIPIcs.ECOOP.2022.7}
}

@InProceedings{atzeni2010,
    author="Atzeni, Paolo
    and Gianforme, Giorgio
    and Toti, Daniele",
    editor="Link, Sebastian
    and Prade, Henri",
    title="Polymorphism in Datalog and Inheritance in a Metamodel",
    booktitle="Foundations of Information and Knowledge Systems",
    year="2010",
    publisher="Springer Berlin Heidelberg",
    address="Berlin, Heidelberg",
    pages="114--132",
    isbn="978-3-642-11829-6"
}

@InProceedings{bernardy2010,
    author="Bernardy, Jean-Philippe and Jansson, Patrik and Claessen, Koen",
    editor="Gordon, Andrew D.",
    title="Testing Polymorphic Properties",
    booktitle="Programming Languages and Systems",
    year="2010",
    publisher="Springer Berlin Heidelberg",
    address="Berlin, Heidelberg",
    pages="125--144",
    isbn="978-3-642-11957-6"
}

@article{somogyi1996,
    title = {The execution algorithm of mercury, an efficient purely declarative logic programming language},
    journal = {The Journal of Logic Programming},
    volume = {29},
    number = {1},
    pages = {17-64},
    year = {1996},
    note = {High-Performance Implementations of Logic Programming Systems},
    issn = {0743-1066},
    doi = {https://doi.org/10.1016/S0743-1066(96)00068-4},
    url = {https://www.sciencedirect.com/science/article/pii/S0743106696000684},
    author = {Zoltan Somogyi and Fergus Henderson and Thomas Conway}
}

@article{vaghanl1994,
    title={The Aditi deductive database system},
    author={Vaghanl, Jayen and Ramamohanarao, Kotagiri and Kemp, David B and Somogyi, Zoltan and Stuckey, Peter J and Leask, Tim S and Harland, James},
    journal={The VLDB Journal},
    volume={3},
    number={2},
    pages={245--288},
    year={1994},
    publisher={Springer},
    doi = { https://doi.org/10.1007/BF01228882 }
}

@article{chiang2023,
    author = {Chiang, David and McDonald, Colin and Shan, Chung-chieh},
    title = {Exact Recursive Probabilistic Programming},
    year = {2023},
    issue_date = {April 2023},
    publisher = {Association for Computing Machinery},
    address = {New York, NY, USA},
    volume = {7},
    number = {OOPSLA1},
    url = {https://doi.org/10.1145/3586050},
    doi = {10.1145/3586050},
    journal = {Proc. ACM Program. Lang.},
    month = apr,
    articleno = {98},
    numpages = {31}
}

@PhdThesis{francislandau2024,
    author={Francis-Landau, Matthew},
    title={Declarative Programming via Term Rewriting},
    school={Johns Hopkins University},
    year={2024},
    url={https://matthewfl.com/papers/mfl-dissertation.pdf}
}

@misc{volkovCommittingBitRelational2025a,
    title = {Committing to the Bit: Relational Programming with Semiring Arrays and {{SAT}} Solving},
    shorttitle = {Committing to the Bit},
    author = {Volkov, Dmitri and Yang, Yafei and Shan, Chung-chieh},
    year = 2025,
    month = sep,
    number = {arXiv:2509.22614},
    eprint = {2509.22614},
    primaryclass = {cs.PL},
    publisher = {arXiv},
    doi = {10.48550/arXiv.2509.22614},
    archiveprefix = {arXiv}
}

@mastersthesis{volkov2026,
    author = {Dmitri Volkov},
    title = {Polymorphic Bottom-Up Weighted Relational Programming},
    school = {Indiana University Bloomington},
    year = {2026}
}

\end{document}